\documentclass[11pt,a4paper,twoside]{article}
\usepackage{color}
\usepackage{ulem}
\usepackage{amsmath}
\usepackage{amsfonts}
\usepackage{rotating}  
\usepackage{graphicx}
\usepackage[T1]{fontenc} 
\usepackage{bm}
\usepackage{hyperref}
\usepackage{version}

\usepackage[english]{babel}
\usepackage{pdflscape}
\usepackage{float}

\numberwithin{equation}{section}
\begin{document}
\vspace{1cm}


\vspace{1cm}

\noindent
{\bf
{\large Holography in flat spacetime: 4D theories and electromagnetic duality on the border
}}

\vspace{.5cm}
\hrule

\vspace{1cm}

\noindent

{\large{Andrea Amoretti$^{a,}$\footnote{\tt andrea.amoretti@ge.infn.it},
Alessandro Braggio$^{b,}$\footnote{\tt alessandro.braggio@spin.cnr.it }, 
Giacomo Caruso$^{a,}$\footnote{\tt giacomo.caruso@ge.infn.it }, 
Nicola Maggiore$^{a,}$\footnote{\tt nicola.maggiore@ge.infn.it },
Nicodemo Magnoli$^{a,}$\footnote{\tt nicodemo.magnoli@ge.infn.it }
\\[1cm]}}

{\small{{}$^a$  Dipartimento di Fisica, Universit\`a di Genova,
via Dodecaneso 33, I-16146, \\Genova, Italy\\and\\INFN, Sezione di Genova\\
\medskip
{}$^b$ CNR-SPIN, Via Dodecaneso 33, 16146, Genova, Italy}}

\vspace{1cm}
\noindent

{\tt Abstract~:}
We consider a free topological model in 5D Euclidean flat spacetime, built from two rank-2 tensor fields. Despite the fact that the bulk of the model does not have any particular physical interpretation, on its 4D planar edge nontrivial gauge field theories are recovered, whose features descend from the gauge and discrete symmetries of the bulk. In particular the 4D dynamics cannot be obtained without imposing a Time Reversal invariance in the bulk. Remarkably, one of the two possible edge models selected by the Time Reversal symmetries displays a true electromagnetic duality, which relates strong and weak coupling regimes. Moreover the same model, when considered {\it on-shell}, coincides with the Maxwell theory, which therefore can be thought of as a 4D boundary theory of a  seemingly harmless 5D topological model.

\vfill\noindent
{\footnotesize 
{\tt Keywords:}
Quantum Field Theory, Duality in Gauge Field Theories, Discrete and Finite Symmetries, Gauge Symmetry, Boundary Quantum Field Theory.
\\
{\tt PACS Nos:} 
03.70.+k	Theory of quantized fields;
11.10.-z	Field theory;
}
\newpage
\begin{small}
\end{small}

\setcounter{footnote}{0}

\section{Introduction}
It is well known that topological field theories acquires local observables only when a considered on a manifold with boundary \cite{Witten:1988hf}. In an actual language, this is the realization of a kind of holographic principle not involving gravity \cite{2013ass}, since the true physical content of a $d+1$ (topological) quantum field theory is encoded in its $d$ boundary. 

There are several relevant examples of this ``correspondence'' : it is realized, for instance, for the 3D Chern-Simons gauge field theory which, when quantized on a manifold with boundary, allows to recover all the states and the representations of the chiral algebra of 2D rational conformal field Theories \cite{Moore:1989yh}. The 2D conserved chiral currents forming a Ka\v{c}-Moody algebra has been explicitly shown to exist for the  Chern-Simons \cite{Emery:1991tf, Blasi:1990pf} and BF \cite{Maggiore:1992ki} 3D topological theories with a planar boundary. This latter has been  introduced according the Symanzik's method \cite{Symanzik:1981wd}, which allows local boundary terms in the action, determined by the general principles of locality, power counting, and a ``decoupling'' condition on the propagators of the theory: the space is divided into a left and a right side, and the propagators between points lying on opposite sides of the boundary must vanish. Later on, the Symanzik's approach to introduce a boundary in quantum field theories has been improved and applied in different situations and in various dimensions \cite{Blasi:2008gt,Blasi:2010gw,Blasi:2011pf,Amoretti:2012hs,Amoretti:2013nv}, with growing attention to condensed matter physics. 
In particular the appearance of topologically ordered materials, such as quantum Hall states \cite{WenZee}, topological insulators \cite{Zhang, Kane} and Weyl semimetals \cite{Weyl}  has
motivated the investigation of topological theories and their peculiar behavior at the boundary. The topological order and the symmetries characterize, again, the physics at the boundary, apart from some non universal constants, which are free parameters for the theory.

An important point of contact between the description of some topological states of matter and topological  field theories with boundary is represented by \cite{Cho:2010rk}, where the 4D BF theory with boundary has been considered to embed a 3D boundary action describing the edge degrees of freedom of the 3D Topological Insulators. In \cite{Amoretti:2012hs} it has been shown that the algebra formed by the conserved edge currents lying on the planar boundary of the 4D BF model can be interpreted as equal time canonical commutation relations, generated by a 3D action which coincides with the one proposed in \cite{Cho:2010rk}. In addition, and remarkably, the boundary condition on the fields of the 4D theory found in \cite{Amoretti:2012hs} is exactly the constraint, called ``duality'' in \cite{Aratyn:1984jz}, which allows to build, at low energy, fermionic fields from bosonic ones, in an analogous way to the fermionization--bosonization procedure which can be exactly carried out in 2D.  This ``duality relation'' is invoked in \cite{Cho:2010rk} to claim the existence, at low energy, of fermionic degrees of freedom, relevant in the description of the 3D Topological Insulators.

The fields of the embedded boundary theory are determined by the gauge symmetry of the embedding bulk theory. In fact, since the full Poincar\'e invariance is anyway lost due to the presence of the boundary, the choice of an axial gauge with axis normal to the boundary appears natural. It is well known that the axial gauge fixing does not completely fix the gauge \cite{Bassetto:1991ue}, so that residual Ward identities remain, one for each gauge symmetry of the bulk theory. The presence of the boundary breaks these Ward identities, and these breakings play the double role of fixing the gauge and of determining the boundary fields. The general rule is that to each $p$ form in the bulk, corresponds a $p-1$-form on the boundary, related one to each other by duality conditions which are the remnant of the boundary conditions on the bulk fields \cite{Amoretti:2012hs,Amoretti:2013nv}.

The aim of this paper is to build nontrivial 4D gauge theories of two interacting gauge fields, following the  flat spacetime holographic principle \cite{2013ass} described above. Therefore, the bulk theory should  be a 5D topological field theory built with two rank-2 tensor fields, say $B_{\mu\nu}$ and $C_{\mu\nu}$, invariant under two gauge transformations, one for each field. This model has also been considered in the context of D-brane models \cite{Maldacena:2001ss,Schwarz:1993vs}. Moreover, in our ``top-down'' approach, we are not interested to what happens at both sides of a planar boundary, as envisaged in the Symanzik's approach, but just on one side. The boundary is then realized by limiting the 5D action by means of a Heaviside step function $\theta(x)$.

 It is interesting to notice that the bulk 5D model which fits our requests turns out to correspond to the one considered recently in \cite{Kravec:2013pua} in order to study its surface states. There, the aim is similar to ours: that is to characterize certain states of matter from properties of their edge states. Although the framework, the motivations and the language of \cite{Kravec:2013pua} are different from ours, there are several intriguing analogies, besides the general aim as we said above. For instance, it is claimed that the edge states are realized locally by the breaking of ``some symmetry'', and are ``protected'' by a discrete symmetry, called ``electromagnetic duality'' involving the two fields. This is exactly what we stated above: the boundary fields and the boundary actions are determined by the broken Ward identities, and we anticipate that it will turn out that the gauge models which we will find on the 4D boundary are identified (and protected) by discrete symmetries in the bulk.

The paper is organized as follows. In Section 2 we write the 5D topological bulk action, with the axial gauge fixing for the two tensor fields $B$ and $C$, to which we add the most general local boundary term, compatible with power counting. In Section 3 the symmetries of the bulk theory are described: the broken residual Ward identities and the discrete symmetries involving the reversal of the $x_0$-coordinate (which by simplicity we call ``Time Reversal''). In Section 4 the most general boundary conditions are derived, classified and discussed. In Section 5 the boundary actions are derived, first by finding out from the broken Ward identities the algebra of the boundary field operators, then translating them in terms of canonical commutation relations between the boundary fields and then finding the most general 4D actions which fit the canonical commutation relations and the boundary conditions found previously and written in terms of boundary fields. Our results are summarized and discussed in the concluding Section 6.

\section{The Action} 

In absence of a boundary, we consider the following bulk action depending on two rank-2 tensor fields $B_{\mu\nu}$ and $C_{\mu\nu}$, built in the 5D flat Euclidean spacetime:
\begin{equation}
S=\int d^5x\ \epsilon_{\mu\nu\rho\sigma\tau}B_{\mu\nu}\partial_\rho C_{\sigma\tau},
\label{noboundaryaction}\end{equation}
which is the most general one invariant under both the following gauge transformations
\begin{eqnarray}
\delta^{(1)} B_{\mu\nu} &=& \partial_\mu c^{(1)}_\nu - \partial_\nu c^{(1)}_\mu\\
\delta^{(1)} C_{\mu\nu} &=& 0
\label{delta1}\end{eqnarray}
and
\begin{eqnarray}
\delta^{(2)} B_{\mu\nu} &=& 0\\
\delta^{(2)} C_{\mu\nu} &=& \partial_\mu c^{(2)}_\nu - \partial_\nu c^{(2)}_\mu,
\label{delta2}\end{eqnarray}
where $c^{(1)}_\mu(x)$ and $c_\mu^{(2)}(x)$ are local gauge parameters. 
We then introduce the boundary at $x_4=0$, implemented by means of  the Heaviside step function $\theta(x_4)$, which changes the usual by parts integration rule into
\begin{equation}
\int d^5x \left[ 
\theta (x_4) 
\epsilon_{\mu\nu\rho\sigma\tau}
\left (
\partial_\rho B_{\mu\nu} C_{\sigma\tau} +  B_{\mu\nu} \partial_\rho C_{\sigma\tau}
\right)
+  \delta(x_4)\epsilon_{ijkl}B_{ij}C_{kj}
\right]
=0,
\label{intpart}\end{equation}
because of the distributional derivative of the $\theta$-function: $\theta^\prime(x)=\delta(x)$,
so that only two of the three terms appearing in \eqref{intpart} are independent. 
Our aim is to study the boundary physics, so we choose to work with the action 
\begin{equation}
S_{bulk} = 
\int d^5x \theta (x_4) 
\epsilon_{\mu\nu\rho\sigma\tau}
\left (
 \partial_\rho B_{\mu\nu} C_{\sigma\tau} +  
k B_{\mu\nu} \partial_\rho C_{\sigma\tau}
\right),
\label{bulkaction}\end{equation}
which depends on one coupling constant $k$, which cannot be reabsorbed by field redefinitions. It must be
\begin{equation}
k\neq1,
\label{kneq1}\end{equation}
because otherwise the action \eqref{bulkaction} would reduce to a pure boundary term, because of \eqref{intpart}.

The notations we adopt in this paper are the following
\begin{eqnarray}
\mu,\nu,\rho,\sigma,\tau &=& 0,1,2,3,4 \\
i,j,k,l &=& 0,1,2,3 \\
\alpha,\beta,\gamma,\delta &=& 1,2,3 \\
\epsilon_{ijkl} &=& \epsilon_{4ijkl} \\
\epsilon_{\alpha\beta\gamma} &=& \epsilon_{40\alpha\beta\gamma}\\
\theta (0) &=& 1\\
x &=& x_\mu = (x_0,x_1,x_2,x_3,x_4)\\
X &=& X_i = (x_0,x_1,x_2,x_3)=(x_0,\vec{X}),
\label{notazioni}\end{eqnarray}
and the canonical mass dimensions of the tensor fields $B$ and $C$ are
\begin{equation}
[B]=[C]=2.
\label{dimensions}\end{equation}
The presence of the boundary $x_4=0$ has as a first consequence that the bulk action is \eqref{bulkaction}, instead of \eqref{noboundaryaction}. In addition, the gauge symmetries of the bulk action are broken by the boundary.

The complete (classical) action is given by
\begin{equation}
S_{tot} = S_{bulk} + S_{gf} + S_J + S_{bd},
\label{totaction}\end{equation}
where
\begin{equation}
S_{gf} = \int d^5x \theta(x_4) ( b_iB_{4i} + d_iC_{4i} ) 
\label{gfaction} \end{equation}
implements the axial gauge choices
\begin{equation}
B_{4i} = C_{4i} =0,
\label{axialgauge}\end{equation}
\begin{equation}
S_{J} = \int d^5x \theta(x_4) ( \frac{1}{2} J^{(B)}_{ij}B_{ij} + 
                                                    \frac{1}{2} J^{(C)}_{ij}C_{ij})
\label{sourceaction}\end{equation}
couples external sources $J^{(B)}$ and $J^{(C)}$ to the tensor fields $B$ and $C$ respectively, and
\begin{eqnarray}
S_{bd} &=& \int d^5x \delta(x_4) ( 
a_1B_{ij}B_{ij} + a_2 \epsilon_{ijkl}B_{ij}B_{kl} + \nonumber \\
&& + a_3C_{ij}C_{ij} + a_4 \epsilon_{ijkl}C_{ij}C_{kl} + a_5 B_{ij}C_{ij}
)\label{bdaction}
\end{eqnarray}
is the most general boundary term, compatible with locality and power counting, depending on five constant real parameters $a_i$, which will turn out to be constrained by the symmetries of the model.
Notice that in \eqref{bdaction} a boundary term of the type $a_6  \delta(x_4) \epsilon_{ijkl}B_{ij}C_{kl}$ has not been included because it can be reabsorbed in \eqref{bulkaction} by means of the integration by parts \eqref{intpart}.

\section{Equations of motion, Ward identities and symmetries} 
From the action \eqref{totaction}, the equations of motion are derived

\begin{eqnarray}
\frac{\delta S_{tot}}{\delta B_{ij}}
&=&
\theta (x_4) [
\frac{1}{2}J^{(B)}_{ij} + (k-1)\epsilon_{ijkl}(\partial_4C_{kl}-2\partial_kC_{4l})
]
\nonumber \\
&&
+ \delta(x_4) [
-\epsilon_{ijkl}C_{kl} + 2a_1 B_{ij} + 2a_2 \epsilon_{ijkl}B_{kl} + a_5C_{ij}]=0
\label{bijeom} \\
\frac{\delta S_{tot}}{\delta B_{4i}}
&=&
\theta (x_4) [
2(k-1)\epsilon_{ijkl}\partial_jC_{kl} + b_i]=0
\label{b4ieom}\\
\frac{\delta S_{tot}}{\delta C_{ij}}
&=&
\theta (x_4) [
\frac{1}{2}J^{(C)}_{ij} + (1-k)\epsilon_{ijkl}(\partial_4B_{kl}-2\partial_kB_{4l})
]
\nonumber \\
&&
+ \delta(x_4) [
-k\epsilon_{ijkl}B_{kl} + 2a_3 C_{ij} + 2a_4 \epsilon_{ijkl}C_{kl} + a_5B_{ij}]=0
\label{cijeom} \\
\frac{\delta S_{tot}}{\delta C_{4i}}
&=&
\theta (x_4) [
2(1-k)\epsilon_{ijkl}\partial_jB_{kl} + d_i]=0,
\label{c4ieom}
\end{eqnarray}
which yield the Ward identities
\begin{eqnarray}
\int_0^{+\infty}dx_4 \partial_j J^{(B)}_{ij} &=&
\left. 2(k-1)\partial_j\tilde{C}_{ij} \right|_{x_4=0}
\label{brokenwijb}\\
\int_0^{+\infty}dx_4 \partial_jJ^{(C)}_{ij} &=&
\left. 2(1-k)\partial_j\tilde{B}_{ij} \right|_{x_4=0},
\label{brokenwijc}
\end{eqnarray}
where we adopted the short-hand notation
\begin{equation}
\tilde{X}_{ij}\equiv\epsilon_{ijkl}X_{kl}.
\end{equation}
It is well known that the axial gauge \eqref{axialgauge} does not completely fix the gauge \cite{Bassetto:1991ue}. The broken Ward identities \eqref{brokenwijb} and \eqref{brokenwijc} are the functional description of the residual gauge invariance (due to the axial gauge choice), broken by the boundary $x_4=0$. From the broken Ward identities \eqref{brokenwijb} and \eqref{brokenwijc}, remembering that $k\neq 1$ \eqref{kneq1}, one immediately sees that Dirichlet boundary conditions for the fields $B$ and $C$ at the boundary $x_4=0$ would imply that also the vanishing of the corresponding l.h.s. This would trivialize the physics on the boundary, which instead is what we are looking for.  This is even more true in our case, since we are considering a $topological$ field theory in the bulk, which lacks of local observables and physical degrees of freedom. It is well known that the only way for a topological field theory to get physical observables is to look what happens on a boundary. Therefore, for our purposes Dirichlet boundary conditions are not interesting, and will not be considered in what follows:
\begin{equation}
\left.B_{ij}\right|_{x_4=0}\neq 0\ \ ;\ \ 
\left.C_{ij}\right|_{x_4=0}\neq 0.
\label{nodirichlet}\end{equation}


Besides the continuum symmetries described by \eqref{brokenwijb} and \eqref{brokenwijc}, the action $S_{bulk}$ \eqref{bulkaction} is also invariant under the following two discrete symmetries, which we call ``Time Reversal'', because they both involve the ``time'' inversion $x_0\rightarrow-x_0$:\footnote{We are aware that calling this discrete symmetry ``Time Reversal'' could be misleading, since in Euclidean spacetime all directions are equivalent. Nonetheless we adopt this nomenclature, as it is widely done in the Literature, having in mind the possible analytic continuation to Minkowski spacetime}
\begin{eqnarray}
T_1B_{04} = + B_{04} &\ & T_1C_{04} = - C_{04} \nonumber\\
T_1B_{4\alpha} = - B_{4\alpha} &\ & T_1C_{4\alpha} = + C_{4\alpha} \nonumber\\
T_1B_{0\alpha} = + B_{0\alpha} &\ & T_1C_{0\alpha} = - C_{0\alpha} \label{t1}\\
T_1B_{\alpha\beta} = - B_{\alpha\beta} &\ & T_1C_{\alpha\beta} = + C_{\alpha\beta} \nonumber
\end{eqnarray}
and
\begin{eqnarray}
T_2B_{04} = - C_{04} &\ & T_2C_{04} = - B_{04} \nonumber\\
T_2B_{4\alpha} = + C_{4\alpha} &\ & T_2C_{4\alpha} = + B_{4\alpha} \nonumber\\
T_2B_{0\alpha} = - C_{0\alpha} &\ & T_2C_{0\alpha} = - B_{0\alpha} \label{t2}\\
T_2B_{\alpha\beta} = + C_{\alpha\beta} &\ & T_2C_{\alpha\beta} = + B_{\alpha\beta}. \nonumber
\end{eqnarray}
One has, indeed
\begin{eqnarray}
T_1S_{bulk} &=& S_{bulk} \label{t1sym} \\
T_2S_{bulk} &=& S_{bulk} + (1+k)\int d^5x\delta(x_4)\epsilon_{ijkl}B_{ij}C_{kl}.
\label{t2sym}
\end{eqnarray}
So $T_2$ is a symmetry of $S_{bulk}$ provided that  $k=-1$.
Imposing the discrete symmetries $T_{1(2)}$ on the boundary action $S_{bd}$ \eqref{bdaction}, yields the following constraints on the parameters $a_i$:
\begin{center}
\begin{equation}
\begin{tabular}{|l|l|l|}
\hline
& $T_1$ &$T_2$\\ \hline
$a_1$&$=a_1$ &$=a_1$ \\ \hline
$a_2$&$=0$ &$=a_2$ \\ \hline
$a_3$&$=a_3$ & $=a_1$ \\ \hline
$a_4$&$=0$ &$ =-a_2$ \\ \hline
$a_5$&$=0$ & $=a_5$ \\ \hline
$k$&$=k$& $=-1$ \\ \hline
\end{tabular}
\label{tabsym}
\end{equation}
\end{center}

\section{Boundary conditions}

The boundary conditions are obtained putting equal to zero the boundary term in the equations of motion \eqref{bijeom} and \eqref{cijeom}: 
\begin{eqnarray}
\left.
-\tilde{C}_{ij} + 2a_1 B_{ij} + 2a_2 \tilde{B}_{ij} + a_5C_{ij}
\right|_{x^4=0} &=& 0
\label{bc1}\\
\left.
-k\tilde{B}_{ij} + 2a_3 C_{ij} + 2a_4 \tilde{C}_{ij} + a_5B_{ij}
\right|_{x^4=0} &=& 0.
\label{bc2}
\end{eqnarray}
The task is to find out which are the parameters $a$ which lead to nonvanishing solutions of the $6 + 6$ equations \eqref{bc1} and \eqref{bc2} for the $6 + 6$ components of the fields $B_{ij}$ and $C_{ij}$. To each solution, it corresponds a boundary condition for the fields $B$ and $C$, which will be crucial for determining the physics on the boundary.

\subsection{Solution imposing $T_1$ }

This corresponds to putting  in \eqref{bc1} and in \eqref{bc2} 
\begin{equation}
a_2=a_4=a_5=0, 
\label{condp1t1}\end{equation}
which therefore become
\begin{eqnarray}
\left.
-\epsilon_{ijkl}C_{kl} + 2a_1 B_{ij} 
\right|_{x^4=0} &=& 0
\label{p1t1bc1}\\
\left.
-k\epsilon_{ijkl}B_{kl} + 2a_3 C_{ij} 
\right|_{x^4=0} &=& 0
\label{p1t1bc2}
\end{eqnarray}

Remembering that Dirichlet boundary conditions are excluded, we observe that the boundary conditions \eqref{p1t1bc1} and \eqref{p1t1bc2} are compatible one with each other if 
\begin{equation}
a_1a_3=k
\label{solp1t1}\end{equation}

The resulting boundary condition is 
\begin{equation}
\left.
B_{ij} - \frac{1}{2a_1} \epsilon_{ijkl}C_{kl} 
\right|_{x^4=0}=0.
\label{sol1bc}\end{equation}
This solution corresponds to the situation studied in \cite{Kravec:2013pua}.

\subsection{Solutions imposing $T_2$ }

This is realized by putting in  \eqref{bc1} and in  \eqref{bc2} 
\begin{equation}
a_3=a_1\ ;\ a_4=-a_2\ ;\ k=-1. 
\label{condp2t2}\end{equation}

The resulting boundary conditions are 
\begin{eqnarray}
\left.
-\epsilon_{ijkl}C_{kl} + 2a_1 B_{ij} + 2a_2 \epsilon_{ijkl}B_{kl} + a_5C_{ij}
\right|_{x^4=0} &=& 0
\label{p2t2bc1}\\
\left.
\epsilon_{ijkl}B_{kl} + 2a_1 C_{ij} - 2a_2 \epsilon_{ijkl}C_{kl} + a_5B_{ij}
\right|_{x^4=0} &=& 0.
\label{p2t2bc2}
\end{eqnarray}
The solutions of the above systems are:
\begin{enumerate}
\item
\begin{eqnarray}
B_{ij} &=& \kappa_1\tilde{B}_{ij} + \kappa_2\tilde{C}_{ij} \label{solt2comp-1}\\
C_{ij} &=& -\kappa_2\tilde{B}_{ij} - \kappa_1\tilde{C}_{ij}
\label{solt2comp-2}\end{eqnarray}
where 
\begin{equation}
\kappa_1 = \frac{4a_1a_2-a_5}{4(1-4a_2^2)}
\ \ ;\ \ 
\kappa_2 = \frac{-2a_1+2a_2a_5}{4(1-4a_2^2)}, 
\label{alpha1beta1}\end{equation}
and $a_1 =\pm \frac{1}{2} \sqrt{16 a_2^2+a_5^2-4 }$, with $16 a_2^2+a_5^2-4\geq 0$.
Notice that the solutions \eqref{solt2comp-1} and \eqref{solt2comp-2} are compatible one with each other because it turns out to hold
\begin{equation}
4(\kappa_1^2-\kappa_2^2)=1.
\label{miracolo}\end{equation}
\item
\begin{equation}
C_{ij}=\pm B_{ij}, 
\label{424}\end{equation}
which is obtained for
\begin{equation}
a_2=\pm\frac{1}{2}\ ;\
a_3=a_1\ ;\
a_4=\mp\frac{1}{2}\ ;\
a_5=\mp 2a_1.
\end{equation}
\item
\begin{eqnarray}
B_{ij} &=& \mp\kappa_1\tilde{B}_{ij} + \kappa_2\tilde{C}_{ij} \label{746}\\
C_{ij} &=& -\kappa_2\tilde{B}_{ij} \pm \kappa_1\tilde{C}_{ij}
\label{747}\end{eqnarray}
where 
\begin{equation}
\kappa_1 = \frac{1+a_1^2}{4a_1}
\ \ ;\ \ 
\kappa_2 = \frac{1-a_1^2}{4a_1}, 
\end{equation}
and, again, $4(\kappa_1^2-\kappa_2^2)=1$, as it should be.
These solutions are analogous to \eqref{solt2comp-1} and \eqref{solt2comp-2}, and are obtained for
\begin{equation}
a_2=\pm\frac{1}{2}\ ;\
a_3=a_1\ ;\
a_4=\mp\frac{1}{2}\ ;\
a_5=\pm 2a_1.
\end{equation}
\end{enumerate}

\subsection{Solutions without imposing discrete symmetries}

Without imposing any of the two discrete $TR$-symmetries \eqref{t1} and \eqref{t2}, the solutions which do not involve Dirichlet boundary conditions are
\begin{enumerate}
\item
\begin{equation}
a_1^{(\pm)}= \pm 2a_2 +
\frac{ (a_5 \pm 2)(a_5 \pm 2k)}{4 (a_3\mp 2 a_4)},
\label{solgen2}\end{equation}
where
\begin{equation}
a_5\pm 2k \neq 0\ ;\
a_3\mp 2a_4\neq 0.
\end{equation}
The boundary conditions are
\begin{equation}
B_{ij}=\mp \frac{1}{2}\tilde{B}_{ij}\ ;\
C_{ij}=\mp \frac{1}{2}\tilde{C}_{ij}\ ;\
C_{ij}=\lambda^{(\pm)}_1 B_{ij},
\label{solgen2sint}\end{equation}
where $\lambda^{(\pm)}_1=\pm\frac{(a_5 \pm  2 k)}{2 (2a_4 \mp  a_3)}$. 
\item
\begin{equation}
a_5=\pm 2k\ \ ;\ \ a_3 = \mp 2 a_4\label{solgen6},
\end{equation}
to which correspond the boundary conditions
\begin{equation}
B_{ij}=\pm \frac{1}{2}\tilde{B}_{ij}\ ;\
C_{ij}=\pm \frac{1}{2}\tilde{C}_{ij}\ ;\
C_{ij}=\lambda^{(\pm)}_2B_{ij},
\label{796}\end{equation}
where $\lambda^{(\pm)}_2=\frac{2a_2\pm a_1}{1-k}$, 
and $2a_2\pm a_1\neq 0$, which otherwise would  imply the forbidden Dirichlet conditions. The case $2a_2\pm a_1=0\ and \ a_5\mp 2k_2=0$ gives   \eqref{solgen2sint}.
\end{enumerate}

\section{Boundary actions}
In this Section we identify the dynamical term of the boundary action using the Ward identities. After that, we will derive the complete 4D action compatible with the boundary conditions found in Section 4.
\subsection{Equal time commutators}

Going {\it on-shell}, that is putting $J=0$ in \eqref{brokenwijb} and in \eqref{brokenwijc}, it is possible to identify the ``potential'' fields which are the correct variables on which the boundary action depends:  

\begin{eqnarray}
\left.\epsilon_{ijkl}\partial_jC_{kl} \right|_{x^4=0} = 0 &\Longrightarrow& 
\left.C_{ij}\right|_{x^4=0} \equiv \partial_i\xi_j(X)-\partial_j\xi_i \label{xi}(X)\\
\left.\epsilon_{ijkl}\partial_jB_{kl} \right|_{x^4=0} = 0 &\Longrightarrow& 
\left.B_{ij}\right|_{x^4=0} \equiv \partial_i\zeta_j(X)-\partial_j\zeta_i, \label{zeta}(X)
\end{eqnarray}
where $X$ has been defined in the notations \eqref{notazioni}.

Deriving the Ward identity \eqref{brokenwijb} with respect to  $J^{(B)}_{mn}(x')$, one gets:
\begin{equation}
(\delta_{\alpha m}\delta_{jn}-\delta_{\alpha n}\delta_{jm})\partial_j\delta^{(4)}(X-X')
=
2(1-k)\epsilon_{\alpha\beta\gamma}\delta(t-t')[C_{\beta\gamma}(X),B_{mn}(X')].
\label{8.4bis}\end{equation}
Putting in \eqref{8.4bis}  $m=\delta, n=\eta$, we get the following equal-time canonical commutation relation:
\begin{equation}
4(1-k)\delta(t-t')[\epsilon_{\alpha\gamma\delta}\partial_\gamma\xi_\delta(X),\zeta_\beta({X'})]
=
\delta_{\alpha\beta}\delta^{(4)}(X-X')
\label{commcan1}\end{equation}
between the 4D canonically conjugate variables
\begin{equation}
q_\alpha(X)\equiv \epsilon_{\alpha\beta\gamma}\partial_\beta\xi_\gamma(X)\ \ ;\ \
p_\beta(X')\equiv4(1-k)\zeta_\beta(X'),
\label{pq1}\end{equation}
in terms of which Eq. \eqref{commcan1} reads
\begin{equation}
[q_\alpha (X),p_\beta (X')] =\delta_{\alpha\beta}\delta^{(4)}(X-X').
\end{equation}
Similarly, deriving the Ward identity \eqref{brokenwijb} with respect to $J^{(C)}_{mn}(x')$ we find
\begin{equation}
\delta(t-t')[C_{\alpha\beta}(X),C_{mn}(X')]=0.
\end{equation}
In terms of the potential $\xi$ and putting $m=\gamma, n=\delta$ one has
\begin{equation}
\delta(t-t')[\epsilon_{\alpha\rho\sigma}\partial_\rho\xi_\sigma(X),\epsilon_{\beta\gamma\delta}\partial_\gamma\xi_\delta(X')]=0,
\label{commcan1.1}\end{equation}
which, according to the identification \eqref{pq1}, corresponds to
\begin{equation}
\delta(t-t')[q_\alpha (X),q_\beta (X')] =0,
\end{equation}
as it should. Observing that $\eqref{brokenwijb} \leftrightarrow \eqref{brokenwijc}$ if $B\leftrightarrow C$ and 
$(k-1)\leftrightarrow (1-k)$, one finds the commutation relation 
\begin{equation}
\delta(t-t')[\zeta_\alpha(X),\zeta_\beta(X')]=0,
\label{commcan1.2}\end{equation}
which, in terms of \eqref{pq1}, corresponds to
\begin{equation}
\delta(t-t')[p_\alpha (X),p_\beta (X')] =0,
\end{equation}
which complete the algebra of the boundary fields.

It is possibile at this point to give a further motivation to the statement according to which the Dirichlet conditions are the only ones which allow a non-trivial boundary physics, whenever an algebra such as that found above is present. Just as one cannot simultaneously impose the initial condition $x$ = $p$ = 0 on a non-relativistic particle, one cannot impose boundary conditions on both members of a pair of canonically conjugate variables such as $B_{0\alpha}$ and $\epsilon_{\alpha\beta\gamma}C_{\beta\gamma}$.

Finally, deriving the Ward identity  \eqref{brokenwijc} with respect to $J^{(C)}_{mn}(x')$, we have:
\begin{eqnarray} 
(\delta_{\alpha m}\delta_{jn}-\delta_{\alpha n}\delta_{jm})\partial_j\delta^{(4)}(X-X')
&=&
2(k-1)\epsilon_{\alpha\beta\gamma}\delta(t-t')[B_{\beta\gamma}(X),C_{mn}(X')],
\nonumber
\end{eqnarray}
where $j=0\rightarrow i=\alpha,k=\beta,l=\gamma$, and we have used \eqref{xi}. The equal time canonical commutation relation is obtained 
putting  $m=\delta, n=\eta$:
\begin{equation}
4(k-1)\delta(t-t')[\epsilon_{\alpha\rho\sigma}\partial_\rho\zeta_\sigma(X),\xi_\beta(X')]
=
\delta_{\alpha\beta}\delta^{(4)}(X-X')
\label{commcan2}\end{equation}
between the canonically conjugate variables
\begin{equation}
q_\alpha(X)\equiv \epsilon_{\alpha\beta\gamma}\partial_\beta\zeta_\gamma(X)\ \ ;\ \
p_\beta(X')\equiv4(k-1)\xi_\beta(X'),
\label{pq2}\end{equation}
in terms of which the \eqref{commcan2} can be written
\begin{equation}
\delta(t-t')[q_\alpha (X),p_\beta (X')] =\delta_{\alpha\beta}\delta^{(4)}(X-X').
\end{equation}
Notice that the identifications \eqref{pq1} and \eqref{pq2} differs one from each other. Nevertheless, they are compatible, in the sense that they both lead to the same action, as we show in the next subsection.

\subsection{Compatibility between \eqref{pq1} and \eqref{pq2}}

The identifications  \eqref{pq1} and \eqref{pq2} are compatible, because they both give rise to the same action.
The Lagrangian density induced by the canonical field variables \eqref{pq1} is, indeed:
\begin{equation}
{\cal L}=p^\alpha\dot{q}_\alpha = 
[4(1-k)\zeta_\alpha (X)]\partial_0[\epsilon_{\alpha\beta\gamma}\partial_\beta\xi_\gamma(X)]
\end{equation}
from which the action is
\begin{equation}
S=\int d^4X\ 4(1-k)\epsilon_{\alpha\beta\gamma}\zeta_\alpha \partial_0\partial_\beta\xi_\gamma.
\label{ax1}\end{equation}
On the other hand, the Lagrangian corresponding to  \eqref{pq2} is:
\begin{equation}
{\cal L}=p^\alpha\dot{q}_\alpha = 
[4(k-1)\xi_\alpha (X)]\partial_0[\epsilon_{\alpha\beta\gamma}\partial_\beta\zeta_\gamma(X)],
\end{equation}
which, integrated, gives rise to the 4D action
\begin{eqnarray}
S &=&
\int d^4X\ 4(k-1)\epsilon_{\alpha\beta\gamma}\xi_\alpha \partial_0\partial_\beta\zeta_\gamma
\nonumber \\ &=&
\int d^4X\ 4(k-1)\epsilon_{\alpha\beta\gamma}\zeta_\gamma \partial_0\partial_\beta\xi_\alpha
\nonumber \\ &=&
\int d^4X\ 4(1-k)\epsilon_{\alpha\beta\gamma}\zeta_\alpha \partial_0\partial_\beta\xi_\gamma.
\label{ax2}\end{eqnarray}
The actions \eqref{ax1} and \eqref{ax2} do indeed coincide.

\subsection{The 4D boundary actions}

The definitions of the two 4D field ``potentials'' $\xi(X)$ in \eqref{xi} and $\zeta(X)$ in \eqref{zeta} induce two Abelian gauge invariance 
\begin{equation}
\delta^{(1)}\xi_i(X)=\partial_i\theta^{(1)} (X)\ \ ;\ \ \delta^{(2)}\zeta_i(X)=\partial_i\theta^{(2)} (X),
\end{equation}
where $\theta^{(1)}(X)$ and $\theta^{(2)}(X)$ are local gauge parameters.

We are looking for a 4D action $S$ depending on two vectorial field $\xi_i(X)$ and $\zeta_i(X)$ which must satisfy the following constraints:
\begin{enumerate}
\item $S$ must not contain terms with time derivatives others than \eqref{ax1}, in order to preserve the canonical commutation relations \eqref{commcan1}, \eqref{commcan1.1} and \eqref{commcan1.2};
\item $S$ must be covariant in the spatial indices $\alpha=1,2,3$;
\item $S$ must be doubly gauge invariant: $\delta^{(1)}S=\delta^{(2)}S=0$;
\item $S$ must be compatible with the boundary/duality conditions we found throughout this paper:  

\begin{enumerate}
\item\label{gruppo1}
\eqref{sol1bc}, \eqref{solt2comp-1}-\eqref{solt2comp-2}, or \eqref{746}-\eqref{747}, which we summarize as:
\begin{eqnarray}
{B}_{ij} &=&  \kappa_1 \tilde{B}_{ij} + \kappa_2 \tilde{C}_{ij} \label{tipo1.1}\\
{C}_{ij} &=&  \kappa_3 \tilde{B}_{ij} - \kappa_1 \tilde{C}_{ij} \label{tipo1.2},
\end{eqnarray}
where, by consistency,
\begin{equation}
4(\kappa_1^2+\kappa_2\kappa_3)=1;
\label{tipo1consistency}\end{equation}
\item\label{gruppo2}
\eqref{solgen2sint} and \eqref{796},  which are of the type 
\begin{eqnarray}
C_{ij} &=& \lambda B_{ij} \label{tipo2}\\
C_{ij} &=& \pm\frac{1}{2} \epsilon_{ijkl}C_{kl} \\
B_{ij} &=& \pm\frac{1}{2} \epsilon_{ijkl}B_{kl};
\end{eqnarray}
\item\label{gruppo3}
\eqref{424}:
\begin{equation}
C_{ij}=\pm B_{ij},
\label{tipo3}\end{equation}
with no (anti)self-duality conditions like in the latter case.
\end{enumerate}

\item $S$ is a gauge theory, for which we chose  the gauge :
\begin{equation}
\xi_0(X)=\zeta_0(X)=0.
\label{gf}\end{equation}
\end{enumerate}

Defining
\begin{equation}
F_{\alpha\beta}(\xi)\equiv\partial_\alpha\xi_\beta(X)-\partial_\beta\xi_\alpha(X)\ \ ;\ \
G_{\alpha\beta}(\zeta)\equiv\partial_\alpha\zeta_\beta(X)-\partial_\beta\zeta_\alpha(X),
\label{defFG}\end{equation}
the most general 4D local action which satisfies the conditions 1. 2. and 3. is:
\begin{equation}
S = \int d^4X\ \left(
\alpha_1\epsilon_{\alpha\beta\gamma}\partial_0\zeta_\alpha\partial_\beta\xi_\gamma 
+ \alpha_2 F^2(\xi) + \alpha_3 G^2(\zeta) + \alpha_4 F_{\alpha\beta}(\xi)G_{\alpha\beta}(\zeta)\right),
\label{testax}\end{equation}
where  $\alpha_1=4(k-1)$, and $\alpha_2,\alpha_3,\alpha_4$ are dimensionless constants that need to be determined in terms of the $a_i$ appearing in \eqref{bdaction}. One immediately sees that terms of lower dimensions are ruled out by the request of compatibility with the duality constraints. Notice that by power counting, the canonical mass dimensions of the potential fields are:
\begin{equation}
[\xi]=[\zeta]=1.
\label{powercount}\end{equation}

From \eqref{testax}, we derive the field equations of motion:

\begin{eqnarray}
\frac{\delta S}{\delta \zeta_\alpha} &=& -\alpha_1\epsilon_{\alpha\beta\gamma}\partial_0\partial_\beta\xi_\gamma 
+ 4\alpha_3(\partial_\alpha\partial\zeta-\partial^2\zeta_\alpha)
+ 2\alpha_4(\partial_\alpha\partial\xi-\partial^2\xi_\alpha)  \label{eomzeta}\\
\frac{\delta S}{\delta \xi_\alpha} &=& 
\alpha_1\epsilon_{\alpha\beta\gamma}\partial_0\partial_\beta\zeta_\gamma
+ 4\alpha_2(\partial_\alpha\partial\xi-\partial^2\xi_\alpha)
+ 2\alpha_4(\partial_\alpha\partial\zeta-\partial^2\zeta_\alpha).
\label{eomxi}\end{eqnarray}

In the Appendix it is shown that  the compatibility between the equations of motion \eqref{eomzeta}-\eqref{eomxi} and the duality conditions of the type  \eqref{tipo1.1} and \eqref{tipo1.2}  is obtained if

\begin{equation}
\alpha_2=-\frac{1}{2}\alpha_1\kappa_2\ \ ;
\alpha_3=\frac{1}{2}\alpha_1\kappa_3\ \ ;
\alpha_4=-\alpha_1\kappa_1.
\label{solalpha}\end{equation}
In details: 
\begin{enumerate}
\item solution \eqref{sol1bc}:
\begin{equation}
\kappa_1=0\ \ ;\ \ \kappa_2=\frac{1}{2a_1}\ \ ;\ \ \kappa_3=\frac{1}{2}a_1
\end{equation}
The 4D boundary action is 
\begin{equation}
S = 4(k-1)\int d^4X\ [
\epsilon_{\alpha\beta\gamma}\partial_0\zeta_\alpha\partial_\beta\xi_\gamma 
-\frac{1}{4}(\frac{1}{a_1}F^2(\xi) - a_1G^2(\zeta))]
\label{bdaction1}\end{equation}
\item solutions \eqref{solt2comp-1}-\eqref{solt2comp-2} and \eqref{746}-\eqref{747}:
 the boundary action is 
\begin{equation}
S = 8\int d^4X\ (
\epsilon_{\alpha\beta\gamma}\partial_0\xi_\alpha\partial_\beta\zeta_\gamma 
+\frac{1}{2}\kappa_2(F^2(\xi) + G^2(\zeta))+\kappa_1F_{\alpha\beta}(\xi)G_{\alpha\beta}(\zeta)).
\label{bdaction2}\end{equation}
\end{enumerate}

Finally, it is readily seen that, following the same procedure described in the Appendix,  the action \eqref{testax} cannot realize the boundary conditions of the type \eqref{tipo2} in terms of equations of motion, as we did previously. In other words, it is not possible to write a 4D action which allows to recover ``on-shell'' the constraints \eqref{tipo2} and \eqref{tipo3}. Neither it is possible an ``off-shell'' realization of those constraints.  In fact, the condition \eqref{tipo2} can be solved in terms of potential fields:
\begin{equation}
\xi_i=\lambda\zeta_i +\partial_i\phi,
\label{xiphi}\end{equation}
where $\phi(X)$ is a scalar field which must be invariant under translations $\phi(X)\rightarrow\phi(X)+c$, in order to preserve gauge invariance on $\xi_i(X)$. If we substitute \eqref{xiphi} into \eqref{ax1} we get zero, and the requested equal time canonical commutators cannot be recovered. 

It is apparent that the discrete symmetries \eqref{t1} and \eqref{t2} play a striking role, since they select the edge dynamics: 4D boundary actions are possible only if one of the two TR symmetries are requested in the bulk. In this sense the edge states are ``protected'' by the discrete symmetries, as remarked in \cite{Kravec:2013pua}. Moreover, and quite remarkably, the action \eqref{bdaction1} displays a true electromagnetic duality, which exchanges the ``electric'' and ``magnetic'' fields, together with the inversion of the coupling constant. We shall come back to this points in the next concluding Section.

\section{Summary and discussion}

In this paper we considered the topological 5D  action
\begin{equation}
S_{bulk} = 
\int d^5x \theta (x_4) 
\epsilon_{\mu\nu\rho\sigma\tau}
\left (
 \partial_\rho B_{\mu\nu} C_{\sigma\tau} +  
k B_{\mu\nu} \partial_\rho C_{\sigma\tau}
\right),
\label{concl1}
\end{equation}
whose physical content  is entirely confined on its 4D boundary, realized here by means of the $\theta$- function appearing in \eqref{concl1}. The adoption of the axial gauge for the two rank-2 tensor fields $B_{\mu\nu}$ and $C_{\mu\nu}$ $and$ the presence of the boundary results in the broken Ward identities
\begin{eqnarray}
\int_0^{+\infty}dx_4 \partial_j J^{(B)}_{ij} &=&
\left. 2(k-1)\partial_j\tilde{C}_{ij} \right|_{x_4=0}
\label{concl2}\\
\int_0^{+\infty}dx_4 \partial_jJ^{(C)}_{ij} &=&
\left. 2(1-k)\partial_j\tilde{B}_{ij} \right|_{x_4=0}.
\label{concl3}
\end{eqnarray}
From \eqref{concl2} and \eqref{concl3}, at vanishing external sources $J$, the 4D boundary fields $\zeta_i(X)$ and $\xi_i(X)$ are readily derived, as vector ``potentials'' of the 5D tensors $B_{\mu\nu}(x)$ and $C_{\mu\nu}(x)$, respectively:
\begin{eqnarray}
\left.\partial_j\tilde{C}_{ij} \right|_{x^4=0} = 0 &\Longrightarrow& 
\left.C_{ij}\right|_{x^4=0} \equiv \partial_i\xi_j(X)-\partial_j\xi_i (X)\label{concl4}\\
\left.\partial_j\tilde{B}_{ij} \right|_{x^4=0} = 0 &\Longrightarrow& 
\left.B_{ij}\right|_{x^4=0} \equiv \partial_i\zeta_j(X)-\partial_j\zeta_i(X),\label{concl5}
\end{eqnarray}
which imply the gauge invariance on the boundary
\begin{equation}
\delta^{(1)}\xi_i(X)=\partial_i\theta^{(1)} (X)\ \ ;\ \ \delta^{(2)}\zeta_i(X)=\partial_i\theta^{(2)} (X).
\label{concl6}\end{equation}

From the broken Ward identities \eqref{concl2} and \eqref{concl3}, the algebra of the conserved currents defined in \eqref{concl4} and \eqref{concl5} is derived, which, written in terms of potential fields $\zeta$ and $\xi$, reads: 
\begin{eqnarray}
[q_\alpha (X),p_\beta (X')] &=&\delta_{\alpha\beta}\delta^{(4)}(X-X') \label{concl7} \\
\delta(t-t')[q_\alpha (X),q_\beta (X')] &=&0 \label{concl8}\\
\delta(t-t')[p_\alpha (X),p_\beta (X')] &=&0, \label{concl9}
\end{eqnarray}
where
\begin{equation}
q_\alpha(X)\equiv \epsilon_{\alpha\beta\gamma}\partial_\beta\xi_\gamma(X)\ \ ;\ \
p_\beta(X')\equiv4(1-k)\zeta_\beta(X').
\label{concl10}\end{equation}
We made the identifications \eqref{concl10} to make apparent the interpretation of the algebra formed by \eqref{concl7},  \eqref{concl8} and \eqref{concl9} as  equal time canonical commutators derived by some 4D action living at the edge of the 5D theory defined by \eqref{concl1}.

The possible boundary conditions are selected by the discrete symmetries \eqref{t1} and \eqref{t2}, which we called ``Time Reversal'', because they both involve the reversal of the (Euclidean) coordinate $x_0$. In fact, if we require Time Reversal in the bulk, the boundary conditions are of the type
\begin{eqnarray}
{B}_{ij} &=&  \kappa_1 \tilde{B}_{ij} + \kappa_2 \tilde{C}_{ij} \label{concl11}\\
{C}_{ij} &=&  \kappa_3 \tilde{B}_{ij} - \kappa_1 \tilde{C}_{ij} \label{concl12},
\end{eqnarray}
where the $\kappa$'s are known functions of the parameters appearing in the boundary term of the $BC$-action \eqref{bdaction}. Otherwise, if no Time Reversal is imposed, the boundary conditions are 
\begin{equation}
B_{ij}=\lambda C_{ij},
\label{concl13}\end{equation}
together with (anti)selfduality conditions on the fields
\begin{equation}
B_{ij}=\pm\frac{1}{2}\tilde{B}_{ij}\ \ ;\ \ 
C_{ij}=\pm\frac{1}{2}\tilde{C}_{ij}.
\label{concl14}\end{equation}

The main results of this paper are the following:
\begin{enumerate}
\item
At the boundary of the topological 5D action \eqref{concl1} it is possible to define a 4D action which is gauge invariant according to \eqref{concl6} and which yields the canonical commutation relations \eqref{concl6}-\eqref{concl9} only if the boundary conditions of the type \eqref{concl11} and \eqref{concl12} are satisfied, {\it i.e.} only if one of the two Time Reversal discrete symmetries \eqref{t1} or \eqref{t2} are respected. In this sense, the edge states of the  model \eqref{concl1} are ``protected'', as guessed in \cite{Kravec:2013pua}.
\item
We showed that the 4D action which respects the boundary conditions \eqref{concl11} and \eqref{concl12}, with $\kappa_3=-\kappa_2$, is
\begin{equation}
S = 8\int d^4X\ (
\epsilon_{\alpha\beta\gamma}\partial_0\xi_\alpha\partial_\beta\zeta_\gamma 
+\frac{1}{2}\kappa_2(F^2(\xi) + G^2(\zeta))+\kappa_1F_{\alpha\beta}(\xi)G_{\alpha\beta}(\zeta)).
\label{concl15}\end{equation}
Notice that the Time Reversal symmetry \eqref{t2} exchanges the $(B,C)$ fields one with each other.
\item
The discrete symmetry \eqref{t1} corresponds to the ordinary Time Reversal symmetry, under which $B$ (hence $\zeta$) behaves like an electric field, and $C$ (hence $\xi$) is magnetic-like. The  4D action 
\begin{equation}
S = 4(k-1)\int d^4X\ [
\epsilon_{\alpha\beta\gamma}\partial_0\zeta_\alpha\partial_\beta\xi_\gamma 
-\frac{1}{4}(\frac{1}{a_1}F^2(\xi) - a_1G^2(\zeta))]
\label{concl16}\end{equation}
is compatible with the boundary conditions \eqref{concl11} and \eqref{concl12}  which, in terms of 4D field strengths, read
\begin{equation}
G_{\alpha\beta}(\zeta)=\partial_\alpha\zeta_\beta-\partial_\beta\zeta_\alpha=\frac{1}{a_1}\epsilon_{\alpha\beta\gamma}\partial_0\xi_\gamma.
\label{concl17}\end{equation}
The second identity coincides with the ``duality'' constraint between the boundary fields $\xi$ and $\zeta$
which is used in \cite{Aratyn:1984jz,Aratyn:1983bg,Amoretti:2013xya} to build, at low energy, fermionic degrees of freedom from bosonic fields, and which therefore allows to claim the presence, on the boundary and at low energy, of fermionic degrees of freedom, as it has been done in \cite{Cho:2010rk} for the 3D Topological Insulators. 

We stress that the boundary actions \eqref{concl15} and \eqref{concl16} depend on the coefficient $a_i$ appearing in \eqref{bdaction}, which are not entirely determined by the symmetries of the bulk theory, as it should, since they encode non-universal information. 

Notice that we are writing the Maxwell equations in terms of two gauge potentials $\zeta$ and $\xi$, related by a duality relation, in a similar way as described in \cite{Bunster:2011qp}. In fact, once the field $\zeta$ is eliminated in favor of $\xi$ through the duality constraint \eqref{concl17}, which corresponds to going {\it on-shell}, the action \eqref{concl16} becomes manifestly 4D covariant, and reads
\begin{equation}
S=\frac{1-k}{a_1}\int d^4X\ F_{ij}(\xi)F_{ij}(\xi).
\label{concl18}\end{equation}
It is a surprising result that the the Maxwell theory described by \eqref{concl18} comes out as the 4D covariant boundary theory of the 5D topological model \eqref{concl1}, which at first glance lacks physical content. 

Finally, and even most remarkably, the action \eqref{concl16}, considered {\it off-shell} as it stands , displays a true electromagnetic duality, because it is invariant under the symmetry
\begin{equation}
\vec{\xi}\leftrightarrow\vec\zeta\ \ ;\ \ a_1 \rightarrow -\frac{1}{a_1},
\end{equation}
which relates the strong (electric) to the weak (magnetic) regime.
\end{enumerate}

{\bf Acknowledgements}\\
We thank the support of INFN Scientific Initiative SFT: ``Statistical Field Theory, Low-Dimensional Systems, Integrable Models and Applications'' and FIRB - ``Futuro in Ricerca 2012'' - Project HybridNanoDev RBFR1236VV.

\appendix\section{ Appendix}
In this Appendix we explicitly study the compatibility between the action \eqref{testax} and the boundary conditions \eqref{tipo1.1} and \eqref{tipo1.2}.
In terms of $\xi$ and $\zeta$, Eq. \eqref{tipo1.1} for $i=0, j=\alpha$ and for $i=\alpha, j=\beta$
gives:
\begin{eqnarray}
\epsilon_{\alpha\beta\gamma}\partial_0\zeta_\alpha &=&
2[\kappa_1(\partial_\beta\zeta_\gamma-\partial_\gamma\zeta_\beta) +
\kappa_2(\partial_\beta\xi_\gamma-\partial_\gamma\xi_\beta)]
\label{dualita10alpha}\\
\partial_\alpha\zeta_\beta-\partial_\beta\zeta_\alpha &=& 
2\epsilon_{\alpha\beta\gamma} 
(\kappa_1\partial_0\zeta_\gamma + \kappa_2\partial_0\xi_\gamma),
\label{dualita1alphabeta}
\end{eqnarray}
where we used the  gauge conditions \eqref{gf}.

Analogously, Eq. \eqref{tipo1.2} can be written
\begin{eqnarray}
\epsilon_{\alpha\beta\gamma}\partial_0\xi_\alpha &=& 
2 [\kappa_3(\partial_\beta\zeta_\gamma-\partial_\gamma\zeta_\beta) -
\kappa_1(\partial_\beta\xi_\gamma-\partial_\gamma\xi_\beta)]
\label{dualita20alpha}\\
\partial_\alpha\xi_\beta-\partial_\beta\xi_\alpha &=& 
2\epsilon_{\alpha\beta\gamma} 
(\kappa_3\partial_0\zeta_\gamma - \kappa_1\partial_0\xi_\gamma).
\label{dualita2alphabeta}
\end{eqnarray}

Compatibility between \eqref{eomxi} and \eqref{dualita10alpha}: taking into account the gauge choice $\xi_0=\zeta_0=0$, the equation of motion \eqref{eomxi} reads:
\begin{equation}
\partial_\beta [
\epsilon_{\alpha\beta\gamma}\partial_0\zeta_\alpha 
+ \frac{4\alpha_2}{\alpha_1}(\partial_\beta\xi_\gamma-\partial_\gamma\xi_\beta)
+ \frac{2\alpha_4}{\alpha_1}(\partial_\beta\zeta_\gamma-\partial_\gamma\zeta_\beta)]=0,
\end{equation}
which is ``compatible'' with \eqref{dualita10alpha} if
\begin{eqnarray}
2\alpha_2 +  \alpha_1\kappa_2&=& 0\label{comp1} \\
\alpha_4 + \alpha_1\kappa_1&=& 0. \label{comp2} 
\end{eqnarray}

Compatibility between \eqref{eomzeta}-\eqref{eomxi} and \eqref{dualita1alphabeta}: 
\begin{equation}
\kappa_2\frac{\delta S}{\delta \zeta_\alpha}-\kappa_1\frac{\delta S}{\delta \xi_\alpha} = 0 
\end{equation}
implies
\begin{eqnarray}
\partial_\beta [
\epsilon_{\alpha\beta\gamma}
(\kappa_1\partial_0\zeta_\gamma + \kappa_2\partial_0\xi_\gamma)
+ (\partial_\alpha\xi_\beta-\partial_\beta\xi_\alpha)
(\frac{4\alpha_2\kappa_1}{\alpha_1} - \frac{2\alpha_4\kappa_2}{\alpha_1})
&&\\
+ (\partial_\alpha\zeta_\beta-\partial_\beta\zeta_\alpha)
(\frac{2\alpha_4\kappa_1}{\alpha_1} - \frac{4\alpha_3\kappa_2}{\alpha_1})]\nonumber
&=&0,
\end{eqnarray}
which is ``compatible'' with \eqref{dualita1alphabeta} if
\begin{eqnarray}
2\alpha_2\kappa_1 -  \alpha_4\kappa_2&=& 0\label{comp3} \\
\frac{4}{\alpha_1}(2\alpha_3\kappa_2-\alpha_4\kappa_1)&=& 1. \label{comp4} 
\end{eqnarray}

Compatibility between \eqref{eomzeta} and \eqref{dualita20alpha}: Eq. \eqref{eomzeta} can be written
\begin{equation}
\partial_\beta [
\epsilon_{\alpha\beta\gamma}\partial_0\xi_\alpha 
- \frac{2\alpha_4}{\alpha_1}(\partial_\beta\xi_\gamma-\partial_\gamma\xi_\beta)
- \frac{4\alpha_3}{\alpha_1}(\partial_\beta\zeta_\gamma-\partial_\gamma\zeta_\beta)]=0
\end{equation}
which is compatible with  \eqref{dualita20alpha} if
\begin{eqnarray}
2\alpha_3 -  \alpha_1\kappa_3&=& 0\label{comp5} \\
\alpha_4 + \alpha_1\kappa_1&=& 0. \label{comp6} 
\end{eqnarray}

Compatibility between \eqref{eomzeta}-\eqref{eomxi} and \eqref{dualita2alphabeta}: 
\begin{equation}
\kappa_1\frac{\delta S}{\delta \zeta_\rho} +\kappa_3\frac{\delta S}{\delta \xi_\rho}= 0 
\end{equation}
implies
\begin{eqnarray}
\partial_\beta [
2\epsilon_{\alpha\beta\gamma}
(-\kappa_3\partial_0\zeta_\gamma + \kappa_1\partial_0\xi_\gamma)
+ (\partial_\alpha\xi_\beta-\partial_\beta\xi_\alpha)
(\frac{8\alpha_2\kappa_3}{\alpha_1} + \frac{4\alpha_4\kappa_1}{\alpha_1})
&&\\
+ (\partial_\alpha\zeta_\beta-\partial_\beta\zeta_\alpha)
(\frac{4\alpha_4\kappa_3}{\alpha_1} + \frac{8\alpha_3\kappa_1}{\alpha_1})]\nonumber
&=&0,
\end{eqnarray}
which is compatible with  \eqref{dualita2alphabeta} if
\begin{eqnarray}
2\alpha_3\kappa_1 +  \alpha_4\kappa_3&=& 0\label{comp7} \\
\frac{4}{\alpha_1}(2\alpha_2\kappa_3+\alpha_4\kappa_1)&=& -1. \label{comp8} 
\end{eqnarray}

Summarizing, the conditions for the compatibility between the equations of motion of the action \eqref{testax} and the duality conditions of the type \eqref{tipo1.1} and \eqref{tipo1.2}, are
\begin{eqnarray}
4(\kappa_1^2+\kappa_2\kappa_3) &=& 1 \label{alphakappa1} \\
2\alpha_2 +  \alpha_1\kappa_2&=& 0\label{alphakappa5} \\
\alpha_4 + \alpha_1\kappa_1&=& 0 \label{alphakappa6} \\
2\alpha_2\kappa_1 -  \alpha_4\kappa_2&=& 0\label{alphakappa7} \\
\frac{4}{\alpha_1}(2\alpha_3\kappa_2-\alpha_4\kappa_1)&=& 1 \label{alphakappa8} \\
2\alpha_3 -  \alpha_1\kappa_3&=& 0\label{alphakappa9} \\
2\alpha_3\kappa_1 +  \alpha_4\kappa_3&=& 0\label{alphakappa11} \\
\frac{4}{\alpha_1}(2\alpha_2\kappa_3+\alpha_4\kappa_1)&=& -1, \label{alphakappa12} 
\end{eqnarray}
which are solved by
\begin{equation}
\alpha_2=-\frac{1}{2}\alpha_1\kappa_2\ \ ;
\alpha_3=\frac{1}{2}\alpha_1\kappa_3\ \ ;
\alpha_4=-\alpha_1\kappa_1.
\end{equation}



\begin{thebibliography}{100}

\bibitem{Witten:1988hf} 
  E.~Witten,
  {\it ``Quantum Field Theory and the Jones Polynomial,''}
  Commun.\ Math.\ Phys.\  {\bf 121}, 351 (1989).

\bibitem{2013ass}
J.~McGreevy, 
{\it Holography with and without gravity},
Lectures held at the ``2013 Arnold Sommerfeld School on Gauge-gravity duality and condensed matter physics'',
\url{http://www.asc.physik.lmu.de/activities/schools/archiv/2013_asc_school/videos_ads_cmt/mcgreevy/index.html}.

\bibitem{Moore:1989yh} 
  G.~W.~Moore and N.~Seiberg,
  {\it ``Taming the Conformal Zoo,''}
  Phys.\ Lett.\ B {\bf 220}, 422 (1989).

\bibitem{Emery:1991tf} 
  S.~Emery and O.~Piguet,
  {\it ``Chern-Simons theory in the axial gauge: Manifold with boundary,''}
  Helv.\ Phys.\ Acta {\bf 64}, 1256 (1991).

\bibitem{Blasi:1990pf} 
  A.~Blasi and R.~Collina,
  {\it ``The Chern-Simons model with boundary: A Cohomological approach,''}
  Int.\ J.\ Mod.\ Phys.\ A {\bf 7}, 3083 (1992).

\bibitem{Maggiore:1992ki} 
  N.~Maggiore and P.~Provero,
  {\it``Chiral current algebras in three-dimensional BF theory with boundary,''}
  Helv.\ Phys.\ Acta {\bf 65}, 993 (1992).
  [hep-th/9203015].

\bibitem{Symanzik:1981wd} 
  K.~Symanzik,
  {\it ``Schrodinger Representation and Casimir Effect in Renormalizable Quantum Field Theory,''}
  Nucl.\ Phys.\ B {\bf 190}, 1 (1981).

\bibitem{Blasi:2008gt} 
  A.~Blasi, D.~Ferraro, N.~Maggiore, N.~Magnoli and M.~Sassetti,
  {\it ``Symanzik's Method Applied To The Fractional Quantum Hall Edge States,''}
  Annalen Phys.\  {\bf 17}, 885 (2008).
  [arXiv:0804.0164 [hep-th]].

\bibitem{Blasi:2010gw} 
  A.~Blasi, N.~Maggiore, N.~Magnoli and S.~Storace,
  {\it ``Maxwell-Chern-Simons Theory With Boundary,''}
  Class.\ Quant.\ Grav.\  {\bf 27}, 165018 (2010)
  [arXiv:1002.3227 [hep-th]].

\bibitem{Blasi:2011pf} 
  A.~Blasi, A.~Braggio, M.~Carrega, D.~Ferraro, N.~Maggiore and N.~Magnoli,
  {\it ``Non-Abelian BF theory for 2+1 dimensional topological states of matter,''}
  New J.\ Phys.\  {\bf 14}, 013060 (2012).
  [arXiv:1106.4641 [cond-mat.mes-hall]].

\bibitem{Amoretti:2012hs} 
  A.~Amoretti, A.~Blasi, N.~Maggiore and N.~Magnoli,
  {\it ``Three-dimensional dynamics of four-dimensional topological BF theory with boundary,''}
  New J.\ Phys.\  {\bf 14}, 113014 (2012).

\bibitem{Amoretti:2013nv} 
  A.~Amoretti, A.~Blasi, G.~Caruso, N.~Maggiore and N.~Magnoli,
  {\it ``Duality and Dimensional Reduction of 5D BF Theory,''}
  Eur.\ Phys.\ J.\ C {\bf 73}, 2461 (2013)
  [arXiv:1301.3688 [hep-th]].

\bibitem{WenZee}
X.~G.~Wen and A.~Zee,
{\it ``A Classification of Abelian quantum Hall states and matrix formulation of topological fluids,''}
  Phys.\ Rev.\ B {\bf 46}, 2290 (1992).

\bibitem{Zhang} 
X.-L.~Qi and S.-C.~Zhang,
{\it ``Topological insulators and superconductors,''}
Rev.\ Mod.\ Phys.\ {\bf 83}, 1057 (2011).

\bibitem{Kane}
M.~Z.~Hasan and C.~L.~Kane,
{\it ``Colloquium: Topological insulators,''}
Rev.\ Mod.\ Phys.\ {\bf 82}, 3045 (2010).

\bibitem{Weyl}
A.~A.~Burkov~and~L~Balents,
{\it ``Weyl Semimetal in a Topological Insulator Multilayer,''}
Phys.\ Rev.\ Lett.\ {\bf 107}, 127205 (2011).

\bibitem{Cho:2010rk} 
  G.~Y.~Cho and J.~E.~Moore,
  {\it ``Topological BF field theory description of topological insulators,''}
  Annals Phys.\  {\bf 326}, 1515 (2011).
  [arXiv:1011.3485 [cond-mat.str-el]].

\bibitem{Aratyn:1984jz} 
  H.~Aratyn,
  {\it ``Fermions From Bosons In (2+1)-dimensions,''}
  Phys.\ Rev.\ D {\bf 28}, 2016 (1983).

\bibitem{Bassetto:1991ue} 
  A.~Bassetto, G.~Nardelli and R.~Soldati,
  {\it ``Yang-Mills theories in algebraic noncovariant gauges: Canonical quantization and renormalization,''}
  Singapore, Singapore: World Scientific (1991) 227 p
  
 \bibitem{Maldacena:2001ss}
 J.~M.~Maldacena, G.~W.~Moore and N.~Seiberg,
  ``D-brane charges in five-brane backgrounds,''
  JHEP {\bf 0110}, 005 (2001).
  [hep-th/0108152].
 
 \bibitem{Schwarz:1993vs}
J.~H.~Schwarz and A.~Sen,
``Duality symmetric actions,''
  Nucl.\ Phys.\ B {\bf 411}, 35 (1994).
  [hep-th/9304154].
  
\bibitem{Kravec:2013pua} 
  S.~M.~Kravec and J.~McGreevy,
  {\it ``A gauge theory generalization of the fermion-doubling theorem,''}
  Phys.\ Rev.\ Lett.\  {\bf 111}, 161603 (2013).
  [arXiv:1306.3992 [hep-th]].

\bibitem{Aratyn:1983bg} 
  H.~Aratyn,
  {\it ``A Bose Representation For The Massless Dirac Field In Four-dimensions,''}
  Nucl.\ Phys.\ B {\bf 227}, 172 (1983).
  
   \bibitem{Amoretti:2013xya} 
  A.~Amoretti, A.~Braggio, G.~Caruso, N.~Maggiore and N.~Magnoli,
  ``3+1D Massless Weyl spinors from bosonic scalar-tensor duality,''
  arXiv:1308.6674 [hep-th].

\bibitem{Bunster:2011qp} 
  C.~Bunster and M.~Henneaux,
  {\it ``The Action for Twisted Self-Duality,''}
  Phys.\ Rev.\ D {\bf 83}, 125015 (2011)
  [arXiv:1103.3621 [hep-th]].


\end{thebibliography}
\end{document}